\documentstyle[12pt,twoside,fleqn,espcrc1]{article}

\input BoxedEPS
\SetRokickiEPSFSpecial  %% for dvips by Tom Rokicki, for VMS
\HideDisplacementBoxes

\newcommand{\AmS}{{\protect\the\textfont2
  A\kern-.1667em\lower.5ex\hbox{M}\kern-.125emS}}
\newcommand{\half}{\frac{1}{2}}

\title{Quark Substructure and Isobar Effects on 
       Deuteron Form-Factors}

\author{E. Lomon\address{Center for Theoretical Physics and
        Laboratory for Nuclear Science\\
        Massachusetts Institute of Technology\\
        Cambridge, MA 02139}%
        \thanks{This work is supported in part by funds provided by  
         the U.S. Department of Energy (D.O.E.) under cooperative
         research agreement \#DF-FC02-94ER40818 and in part by KEK funds for
Foreign Visiting Scientists.\quad MIT CTP\#2949 \hfil\break
Present address: Theory Group, KEK--Tanashi branch, Tokyo}}

\begin{document}
% typeset front matter
\maketitle

\begin{abstract}
Elastic ed scattering, with deuteron polarization, up to high momentum
transfer provides detailed information on the deuteron wave function. This
determines the range dependence of the orbital and spin components of the
one- and two-body currents, restricting contributions of isobar and
meson-exchange currents and of quark/gluon degrees of freedom, as well as
the nucleon component. The $R$-matrix boundary condition model combines all
these effects, predicting nucleon-nucleon reactions and the deuteron
form-factors simultaneously. A brief description of the model is followed by a
comparison of its results with data, emphasizing the restrictions placed on the
model by ed elastic form-factors.
\end{abstract}

\section{INTRODUCTION}

There is now elastic electron-deuteron scattering data determining the electric
form-factor $A(q^2)$ up to $6~($GeV/$c)^2$\cite{JLAB-A}, the magnetic
form-factor $B(q^2)$ up to $2.8~($GeV/$c)^2$\cite{Arnold87}, and the
tensor-polarization form-factor $t_{20}(q^2)$ up to
$1.8~($GeV/$c)^2$\cite{JLAB-t}.  These data restrict orbital and spin
components of the deuteron wave function at scales as small as 0.2 fm, at
which distance quark degrees-of-freedom (d.o.f.), isobar components and
meson-exchange currents all have a significant role.

The $R$-matrix boundary condition method\cite{Paul86,Pedro87} provides a
hybrid quark/gluon and hadron model, incorporating all the above
contributions.  Only a few of the parameters are not predetermined by data
independent of the nucleon-nucleon (NN) interaction and symmetry
requirements.  The remaining few are almost all determined by NN scattering
data.  Essentially one parameter is free to determine the behavior of the three
independent elastic electron-deuteron form-factors (edff) over the large range
of momentum-transfers, $q$.  This parameter determines the relative amount
of $\Delta\Delta (^7D_1)$ and $\Delta\Delta (^3D_1)$ in the deuteron, which
profoundly affects the $q$ dependence of the spin and convective
currents\cite{Sitar}.  The NN scattering is not sensitive to the ratio, but only
to the sum.

Following a review of the $R$-matrix method and its application to the NN
system, three specific models for the $I=0$, $J^p=I^+$ sector, of different levels
of completeness, will be compared with the NN scattering and edff data.  From
these one can extrapolate to a model that represents all the data.

\section{THE $R$-MATRIX BOUNDARY CONDITION MODEL}

At high momentum-transfer (short range) the running coupling constant of
QCD is small, permitting a perturbative description in terms of current quarks
and gluons (asymptotic freedom).  At low momentum-transfer (long range)
nonperturbative effects produce clustering into color singlet hadrons
(confinement).  The transition between these extremes has been shown to
occur over a small range of the running coupling constant\cite{Creutz}, and
therefore over a short distance.  The $R$-matrix method\cite{Wigner} is well
suited to this situation in which two regions, each well represented by a
different approximate Hamiltonian, have their wave functions connected by a
boundary condition at the separating surface.

For the QCD application, in which confinement requires the quark wave
function to be small at the transition boundary, the suitable form of the
$R$-matrix equation is\cite{Paul86,Pedro87}
\begin{equation}
r_0 \Bigl( \frac{\partial \psi_\alpha (r,W)}{\partial r} \Bigr)_{r_0} = 
\sum_\beta f_{\alpha\beta}(W) \psi_\beta (r_0,W)
\label{eq:1}
\end{equation}
with 
\begin{equation}
f_{\alpha\beta}(W) = f^0_{\alpha\beta} + \sum_i
\frac{\rho^i_{\alpha\beta}}{W-W_i}
\label{eq:2}
\end{equation}
in which $\psi_\alpha$ is the exterior wave function for hadron-pair channel
$\alpha$, $W$ is the total energy, the poles $W_i$ are the energies of a
complete set of internal states vanishing at $r_0$, and the residues
$\rho^i_{\alpha\beta}$ are given by absolute squares of the internal wave
function derivatives at the boundaries.  Thus the $f_{\alpha\beta}(W)$ are
meromorphic functions with real poles of positive residue.  The residues can be
expressed as
\begin{equation}
\rho^i_{\alpha\beta} = -r_0 \frac{\partial W_i}{\partial r_o} \xi^i_\alpha
\xi^i_\beta
\label{eq:3}
\end{equation}
where the fractional parentage coefficients $\xi^i_\alpha$ are geometric
coefficients expressed in terms of Clebsch-Gordon coefficients of the
spin/flavor/color space of the given quark configuration.  Therefore only the
$f^0_{\alpha\beta}$, representing the effective constant of distant poles and
the pole at infinity, are free parameters.

The hadrons at $r>r_0$ interact via hadron exchange potentials, as given by
known hadron masses and coupling constants fixed by independent
experiments or symmetry conditions.

The separation radius, $r_0$, must be within the range of asymptotic freedom
($\le 0.85$ of the equilibrium radius of the interior bag model) because of the
sensitivity of $\rho^i_{\alpha\beta}$ to the derivatives at $r_0$ of the internal
wave function.  Eq.~(\ref{eq:1}) also requires that $r_0$ satisfy $\psi_2 (r_0,
W_i(r_0))=0$.  Using the low energy data, the latter condition fixes the value of
$r_0$ with some precision, giving a value consistent with the first condition
for the Cloudy bag model, but not for the MIT bag model\cite{Paul86,Pedro87},
ruling out the latter for the multi-hadron domain.  The Cloudy bag model
determines $r_0=1.05$~fm.  The model then gives a good detailed fit to NN
data for $T_{\rm Lab} \le 0.8$~GeV\cite{LL} and also is consistent with some
evidence for the lowest $I=1$ exotic resonance, $J^p=0^+$\cite{Ball} at 2.70
GeV.  These resonances, produced near the $f$-poles at $W_i$ correspond to
multi-quark configurations other than the minimal $q\bar{q}$ and $q^3$
configurations.  The lowest in the NN system is the $I=0$, $J^p=1^+$ at 2.63 GeV.

\section{DEUTERON PREDICTIONS}

\subsection{Interior Wave Function}

The first $f$-pole in the NN system, corresponding to the $[q(1s_\half)]^6$
configuration, is in the $I=0$, $J^p=1^+$ state, 0.76 GeV above the deuteron
mass.  As the width of the exotic resonance is only 0.03 GeV, the
$f_{\alpha\beta}$ are nearly constant.  It has been shown\cite{FL} that the
interior wave function vanishes for constant $f$, and the actual probability of
being in the interior has been estimated to be $\le 0.004$\cite{Sitar}.  This
implies a large ``hole'' in the deuteron wave function, which has long been
noted as a property of the edff and is also embodied in the effective ``hard
core'' of NN scattering.  In our model this apparent repulsion arises from the
rapid change in the effective d.o.f. at $r_0$.  However at the energy $W_i$
there is ``matching'' of the interior and exterior wave functions, resulting in a
substantial interior probability and a resonance.

The fact that $r_0$ is 2-3 times larger than the cores of repulsive core models
is compensated by the discontinuous increase of wave functions at $r_0$, so
that the experimental effective range parameters can be produced by both
types of core effect.  However the models differ at higher NN scattering
energies for large $q$ edff.

\subsection{Exterior Wave Function}

In the $I=0$, $J^p=1^+$ system, the NN $(^3S_1)$ and NN $(^3D_1)$ states are
coupled to each other and to isobar channels by meson exchange potentials as
well as the $f$-matrix.  Because of their low threshold mass and strong tensor
coupling, the $\Delta\Delta (^3S_1)$, $\Delta\Delta (^3D_1)$ and $\Delta\Delta
(^3D_1)$ channels are most important and are included in all our models.    NN
$(^3S_1)$ and  $\Delta\Delta (^3S_1)$ states have nonvanishing $\xi^i_\alpha$
with the $[q(1s_\half)]^6$ quark configuration, contributing to the $f$-pole
residue.  The NN$^* (1440) (^3S_1)$ channel also has a low threshold, and
with its large width is of next importance to the deuteron.  But over the
energy range which includes the first exotics, other channels are also of some
importance and the $NS_{11} (1535) (^1P_1)$, $NS_{11} (1650) (^1P_1)$ and
$\Delta S_{31} (1620) (^3P_1)$ are included in our recent work.  These
channels modify the best choice of $f^0_{\alpha\beta}$ for the lower
threshold channels, but have negligible components in the deuteron.

\subsection{Determining the \protect\boldmath$f^0_{\alpha\beta}$ for Three Models}

The $f^0_{\alpha\beta}$ are sharply restricted by fitting the NN scattering
data, which in the deuteron sector requires a fit to the $np(^3S_1-^3D_1)$
phase parameters $\delta$ and $\eta (^3S_1)$,  $\delta$ and $\eta (^3D_1)$
and $\epsilon_1$ for $T_{\rm Lab} \le 0.8$~GeV.  For the models $C'$ and
$D'$\cite{Green}, which have only NN and $\Delta\Delta$ channels, the lower
energy behavior of the $\delta$'s and $\epsilon_1$ determine the NN sector
$f_{\alpha\beta}$, while the energy dependence for 0.4 GeV $<T_{\rm Lab} <$
0.8 GeV fixes those coupling the NN and $\Delta\Delta (^3S_1)$ sectors, and the
NN to the sum of $\Delta\Delta (^3D_1)$ and $\Delta\Delta (^7D_1)$ sectors. 
The last two have the same threshold behavior, affecting the elastic
scattering in the same way.  Only detailed $\Delta$-production data could
separate them, so the ratio is free to adjust to the edff.  The magnetic form
factor $q$-dependence is very sensitive to the ratio because of the opposite
spin and convection currents of these states\cite{Sitar}.

The model $C'$ did not consider quark configurations\cite{LF}.  Without
$f$-poles the best choice of separation radius was $r_0=0.74$~fm.  Model $D'$
included the lowest $f$-pole with $r_0=1.05$~fm as discussed above.  As
shown previously\cite{Green}, $C'$ and $D'$ give equivalent fits to the
$\delta$'s while case $D'$ is a better fit to $\epsilon_1$.
\vspace{-1.5pc}

\begin{figure}[ht]
\centerline{\BoxedEPSF{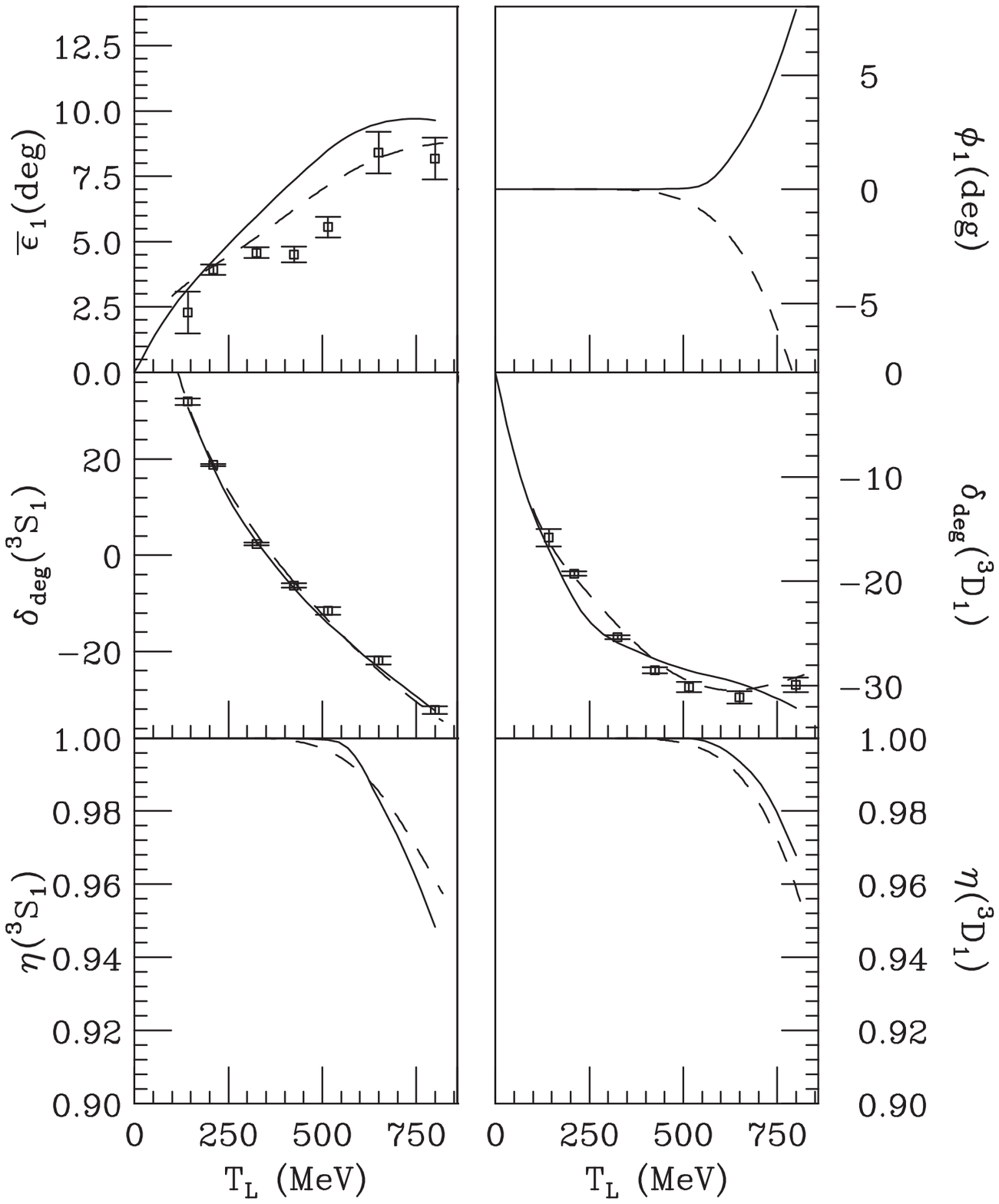 scaled
566}\quad\raise30ex\hbox{\parbox[t]{2.5in}{\quad When the other isobar
channels are included, the $NN^* (1440)$, because of its larger width, modifies the
energy dependence and increases the inelasticity.  This reduces the required coupling
to the
$\Delta\Delta$ channels.  This model $E$ results in a better fit to
$\bar\epsilon_1, \delta (^3D_1)$ and to the
$\eta$'s (Fig.~1).\vspace{10ex}
\caption{The phase parameters for
$I=0$,
$J^P=1^+$,
$np$ scattering. Model~$E$ (\emph{solid curves}); SAID SP00 phase parameters
(\emph{dashed curves}); Bugg 1990--1991 phase parameters (\emph{squares}).}}}}
\label{fig:1}
\end{figure}

\subsection{The EDFF Predictions}

In previous work\cite{Green}, the edff for models $C'$ and $D'$ were
calculated with the nonrelativistic, coupled channel impulse approximation
(IA) and the meson-exchange current (MEC) terms of $\pi$, $\rho$, and
$\omega$, ``pair'' corrections and the $\rho \pi \gamma$ term to first
relativistic order.  For the IA the isobar form factors are assumed
proportional to the nucleon electromagnetic form-factors.  In all cases the MEC
corrections to the isobar channels are neglected.  Both H\"ohler et al.\cite{HO}
(HO) and Gari-Kr\"umpelmann\cite{GK} (GK) nucleon form factors were used.
 The results (Figs.~3-6 of \cite{Green}) are seen to be a good to
$A(q^2)$ for the HO choice and to $B(q^2)$ for the GK choice.  This is not
inconsistent as $A(q^2)$ is dominated by the nucleon electric form-factor and
$B(q^2)$ by the nucleon magnetic form-factor.  The $t_{20}(q^2)$ predictions
were consistent with the very low $q$ experimental results available  at the
time.

Here we present, versus the extended data range of the edff, the results of
models $C'$, $D'$ and $E$ where the first order relativistic correction has been
added to the impulse approximation and the second order relativistic
corrections have been included in the MEC\cite{Blun,Aren}.  For model $E$ the
ratio of $\Delta\Delta(^7D_1)$ to $\Delta\Delta(^3D_1)$ coupling to the NN
sector was guided by the $C'$ and $D'$ model ratios.  It has not yet been
optimized to the data.  Also the balance of $\Delta\Delta$ and $NN^* (1440)$
coupling to NN and the value of the $N^* (1440)$ magnetic moment, unknown
from independent data, have not been varied.

Table~\ref{tab:1} shows the results of model $E$ for the static properties of the
deuteron.  For models $C'$ and $D'$ the results are as stated in \cite{Green}
except for a small relativistic change in $Q_{\rm deut}$.\vspace{-1pc}

\begin{table}[hbt]
% -----------------------------------------------------
% adapted from TeX book, p. 241
\newlength{\digitwidth} \settowidth{\digitwidth}{\rm 0}
%\catcode`?=\active \def?{\kern\digitwidth}
% -----------------------------------------------------
\caption{Static deuteron properties of Model $E$}\label{tab:1}
\begin{tabular*}{\textwidth}{@{}lcccccccc}
&$BE$(MeV) & $P_D(\%)$ &$P_{\Delta 5}(\%)$ &$P_{\Delta 3}(\%)$&$P_{\Delta 7}(\%)$&
$P_{N^*}$(\%)\rlap{$^{\rm a}$}&$Q(fm^2)$&$\mu_D(\mu_{?})$ \\ 
\hline
Model $E$ & 2.2247 & 5.21 & .006 & 3.24 &
1.79 & 0.71 & .273 & .860 \\ Exp $\ell$ & 2.2246 &&&&&&.286 & .857\\
\hline
\multicolumn{9}{@{}p{\textwidth}}{\small $^{\rm a}$The higher mass channels have neglibible
probability.}
\end{tabular*}\vspace{-1pc}
\end{table}

\begin{figure}[bt]
\centerline{\BoxedEPSF{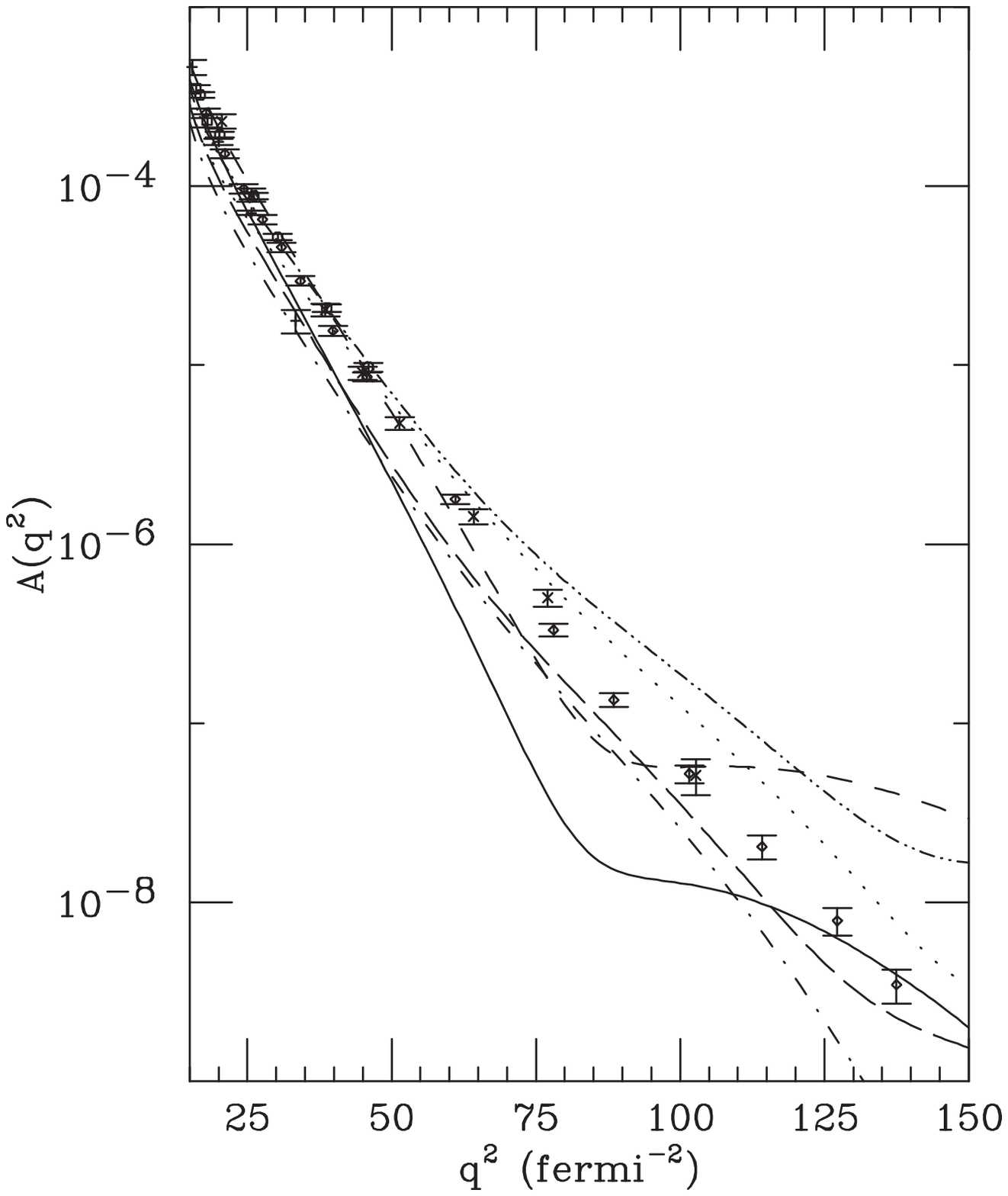 scaled
500}\qquad\BoxedEPSF{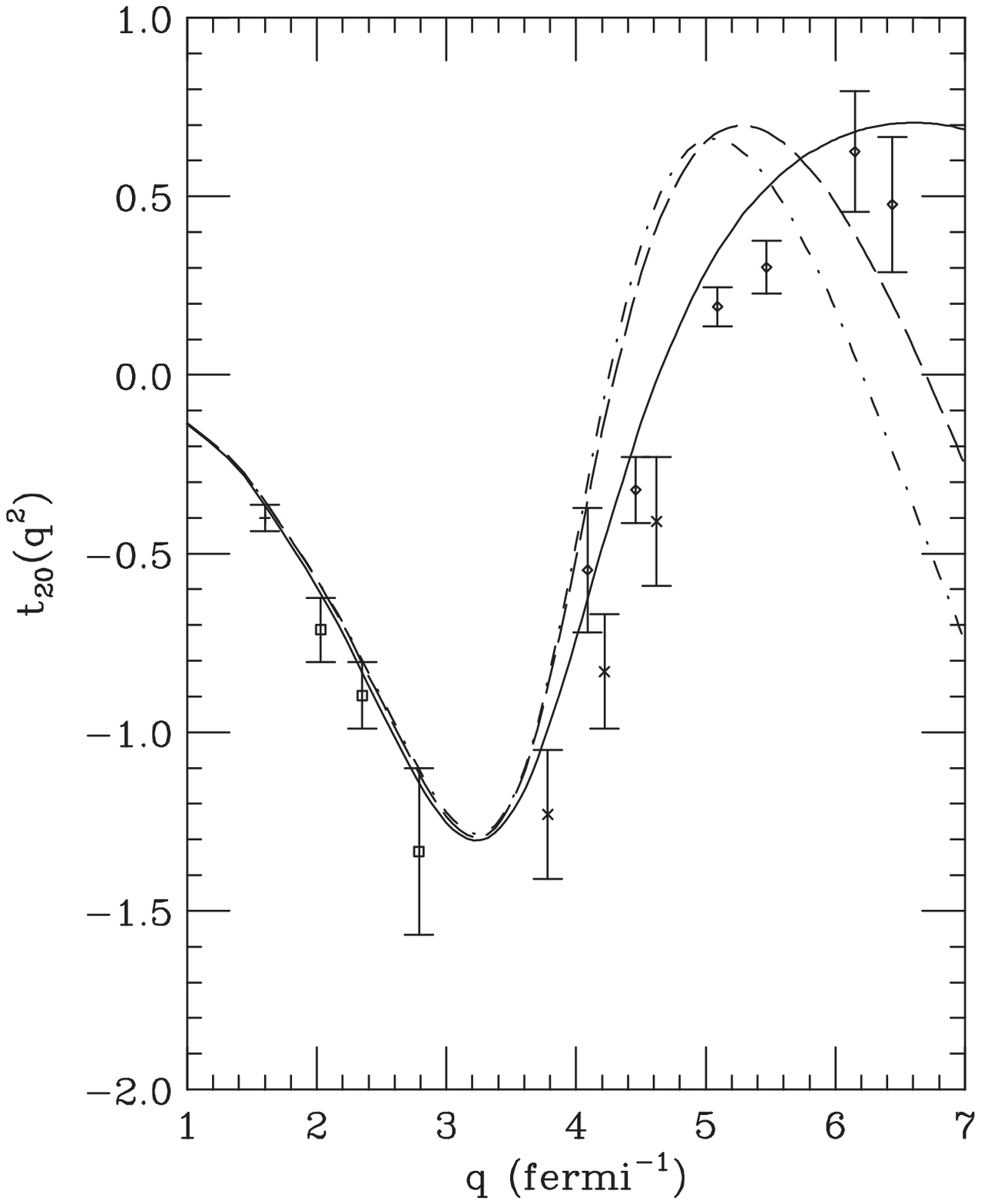 scaled 500}\hfill}
\smallskip
\centerline{\hfill Fig.~2\hfill\hfill
 Fig.3\qquad\hfill}\vspace{-1pc}
\caption{$A(q^2)$: Data points are described in Ref.\cite{JLAB-A}. Model $C'$
(HO) (\emph{solid line}); model $C'$ (GK)  (\emph{dash-dash}); model $D'$
(HO)  (\emph{dash-dot}); model $D'$ (GK) (\emph{dot-dot}); model $E$  (HO)
(\emph{long dashes}); model $E$ (GK) (\emph{dash-dot-dot}).}
\label{fig:2}\vspace{-2pc}
\caption{$t_{20}(q)$: Data points described in Ref.~\protect\cite{JLAB-t}. Model $C'$
(\emph{solid line});  model $D'$  (\emph{dash-dot});  model $E$  (\emph{long
dashes}). The dependence on nucleon emff (HO or GK) is negligible.}\vspace{-1pc}
\label{fig:3}
\end{figure}

$A(q^2)$ is shown in Fig.~2.  It is seen that model $E$ with either choice of
nucleon form-factors fits the data reasonably well for $q^2<2.5$ (GeV/$c$)$^2$,
but is only large enough for larger $q^2$ when the GK choice is made.  Models
$C'$ and $D'$ on the other hand are better with the HO choice for $q^2 <2.5$
(GeV/$c^2$), but at 6 (GeV/$c^2$) only model $C$ with GK is large enough. 

$t_{20}(q^2)$ is shown in Fig.~3.  For the momentum transfer range there is
negligible sensitivity to the choice of nucleon form-factors, as these cancel in
the ratio of quadrupole to monopole electric amplitudes which dominates 
$t_{20}$.  The result is however very sensitive to the model used.  The simple
constant $f$-matrix model, $C'$, gives a good fit over the whole range of $q$. 
Model $D'$ puts the maximum of $t_{20}$ at much too small a momentum
transfer.  This is related to the large amplitude ofthe $L=2$, $\Delta\Delta$
states in this model. Model $E$, with intermediate $\Delta\Delta$ components,
has the maximum of $t_{20}$ between that of models $C'$ and $D'$.

The $B(q^2)$ for models $C'$ and $D'$ is similar to that of \cite{Green} with
minima at slightly smaller $q$.  However the minimum of $B(q^2)$ for model
$E$ is at much too small a value $(q^2=1.3$~(GeV/$c)^2$).

\section{CONCLUSIONS}

The $R$-matrix boundary condition model $E$, with all the relevant isobar
channels, reproduces the $np(^3S_1-^3D_1)$ scattering phases well up to
$T_{\rm Lab} \le 0.8$~GeV.  It also reproduces very well the static properties
of the deuteron and $A(q^2)$ for $q^2 \le 6$~(GeV/$c)^2$.  It does not fit
$t_{20}(q^2)$ as well as the simpler model $C'$ (although it is better than
model $D'$), and has the first minimum of $B(q^2)$ at much too small a value. 
To correct the full model ($E$) for $B(q^2)$, the ratio of the $\Delta\Delta
(^7D_1)$ to $\Delta\Delta (^3D_1)$ couplings to the NN sector needs to be
varied.  That, and perhaps a further substitution of $N^* (1440)$ coupling for
$\Delta\Delta$ coupling to the NN sector may also correct the fit to the
maximum of $t_{20}(q^2)$.

%\begin{table}\caption{Static deuteron properties of Model $E$}
%\begin{tabular*}{\textwidth}{lcccccccc}
%\hline
%&$BE$(MeV) & $P_D(\%)$ &
%$P_{\Delta 5}(\%)$ &  $P_{\Delta 3}(\%)$ &
%$P_{\Delta 7}(\%)$ & $P_{N^*}(\%)$ \footnote{} & $Q(fm^2)$ &
%$\mu_D(\mu_N)$ \\
%Model $E$ & 2.2247 & 5.21 & .006 & 3.24 & 1.79 & 0.71 & .273 & .860 \\
%Exp $\ell$ & 2.2246 &&&&&&.286 & .857
%\footnotetext{The higher mass channels have neglibible probability}
%\end{tabular*}
%\end{table}

\end{document}